\documentclass[11pt,runningheads,letterpaper]{llncs}
\pdfoutput=1

\newcommand{\longversion}[1]{#1}
\newcommand{\shortversion}[1]{}
\usepackage{amsmath,amssymb,amstext,amsfonts} 
\usepackage[latin1]{inputenc} %
\usepackage{graphicx}
\usepackage{xspace}
\usepackage{boxedminipage}

\usepackage{vmargin}
\setmarginsrb{1.35in}{1.00in}{1.35in}{1.70in}{0.5cm}{0.5cm}{0cm}{0cm}

\spnewtheorem{boldclaim}{Claim}{\bfseries}{\itshape}

\newcommand{\set}[1]{\{ #1 \}}

\newcommand{\pbDef}[4]{%
\noindent
\begin{center}
\begin{boxedminipage}{0.99 \textwidth}
#1
\smallskip\\
\begin{tabular}{l p{0.78 \textwidth}}
Input: & #2\\
Parameter: & #3\\
Question: & #4
\end{tabular}
\end{boxedminipage}
\end{center}
}

\newcommand{\ILPF}{\textsc{Integer Linear Programming Feasibility}\xspace}
\newcommand{\SubsetSum}{\textit{va\-ri\-e\-ty}-\textsc{Subset Sum}\xspace}
\newcommand{\Partition}{\textit{va\-ri\-e\-ty}-\textsc{Partition}\xspace}
\newcommand{\NumericalTDMatching}{\textit{va\-ri\-e\-ty}-\textsc{Numerical 3-Di\-men\-sio\-nal Matching}\xspace}
\newcommand{\NumericalMatchingTargetSums}{\textit{va\-ri\-e\-ty}-\textsc{Numerical Matching with Target Sums}\xspace}
\newcommand{\ThreePartition}{\textit{va\-ri\-e\-ty}-\textsc{3-Partition}\xspace}
\newcommand{\ExistsWordMM}{\textit{va\-ri\-e\-ty}-\textsc{Exists Word Mealy Machine}\xspace}
\newcommand{\GivenWordMM}{\textit{va\-ri\-e\-ty}-\textsc{Given Word Mealy Machine}\xspace}
\newcommand{\MCC}{\textsc{Mul\-ti\-co\-lored Clique}\xspace}

\newcommand{\sILPF}{\textsc{ILPF}\xspace}
\newcommand{\sSubsetSum}{\textit{var}-\textsc{SubSum}\xspace}
\newcommand{\sPartition}{\textit{var}-\textsc{Part}\xspace}
\newcommand{\sNumericalTDMatching}{\textit{var}-\textsc{Num3-DM}\xspace}
\newcommand{\sNumericalMatchingTargetSums}{\textit{var}-\textsc{NMTS}\xspace}
\newcommand{\sThreePartition}{\textit{var}-\textsc{3-Part}\xspace}
\newcommand{\sExistsWordMM}{\textit{var}-\textsc{EWMM}\xspace}
\newcommand{\sGivenWordMM}{\textit{var}-\textsc{GWMM}\xspace}
\newcommand{\sMCC}{\textsc{MCC}\xspace}

\DeclareMathOperator*{\gap}{\mathsf{gap}}
\DeclareMathOperator*{\edge}{\mathsf{edge}}
\newcommand{\letter}[1]{\langle #1 \rangle}

\newcommand{\fpt}{fixed-pa\-ra\-me\-ter trac\-ta\-ble\xspace}
\newcommand{\fptred}{parameterized reduction\xspace}
\newcommand{\deterministic}{deterministic\xspace}
\newcommand{\nondet}{non-{\deterministic}\xspace}

\newcommand{\FPT}{\text{\normalfont FPT}}
\newcommand{\XP}{\text{\normalfont XP}}
\newcommand{\W}[1][xxxx]{\text{\normalfont W}[#1]}

\title{Parameterizing by the Number of Numbers\thanks{%
A preliminary version of this paper appeared in the proceedings of IPEC 2010 \cite{FellowsGR10}.
M.R.F. and F.A.R. acknowledge support from the Australian Research Council.
S.G. acknowledges partial support from the European Research Council (COMPLEX REASON, 239962),
from Conicyt Chile (Basal-CMM), and from the Australian Research Council.}}

\author{Michael R. Fellows\inst{1}
 \and Serge Gaspers\inst{2}
 \and Frances A. Rosamond\inst{1}
} 
 
\institute{ 
 School of Engineering and IT,
Charles Darwin University,
NT 0909,
Australia.
 \texttt{\{michael.fellows,frances.rosamond\}@cdu.edu.au}
 \and 
 Institute of Information Systems, Vienna University of Technology, Vienna, Austria.
 \texttt{gaspers@kr.tuwien.ac.at}
}

\date{}

\begin{document}

\maketitle

\begin{abstract}
The usefulness of parameterized algorithmics has often depended on what Niedermeier has called ``the art of problem parameterization''.  In this paper we introduce and explore a novel but
general form of parameterization: {\it the number of numbers}.
Several classic numerical problems, such as \textsc{Subset Sum}, \textsc{Partition}, 3-\textsc{Par\-ti\-tion}, \textsc{Numerical 3-Dimensional Matching}, and \textsc{Numerical Matching with Target Sums}, have multisets of integers as input.  We initiate the study of parameterizing these problems by the number of distinct integers in the input.  We rely on an FPT result for \ILPF to show that all the above-mentioned problems are \fpt when parameterized in this way. 
In various applied settings, problem inputs often consist in part of multisets of integers or multisets of weighted objects (such as edges in a graph, or jobs to be scheduled). 
Such number-of-numbers parameterized problems often reduce to subproblems about transition
systems of various kinds, parameterized by the size of the system description. 
We consider several core problems of this kind relevant to number-of-numbers parameterization. 
Our main hardness result considers the problem: given a \nondet Mealy machine $M$ (a finite state automaton outputting a letter on each transition), an input word $x$, and a census requirement $c$ for the output word specifying how many times each letter of the output alphabet should be written, decide whether there exists a computation of $M$ reading $x$ that outputs a word $y$ that meets the requirement $c$.  We show that this problem is hard for $W[1]$. If the question is whether there exists an input word $x$ such that a computation of $M$ on $x$ outputs a word that meets $c$, the problem becomes \fpt.
\end{abstract}

\section{Introduction}

Parameterized complexity and algorithmics has been developing for more than twenty years.  Some important progress of the field has depended on what Niedermeier has called ``the art of problem parameterization'' 
(see Chapter 5 of his monograph~\cite{Niedermeier06}).  For example, it was Cristina Bazgan
who first suggested that the parameter might be $k = 1 / \epsilon$ in the study of the complexity of
approximation, leading eventually to the study of EPTASs \cite{Bazgan95}. 

Here we explore, for the first time (to our knowledge), a parameterization that
seems widely relevant: {\it the number of numbers}.  Many problems take as input information that consists (in part) of multisets of integers or multisets of weighted objects, such as weighted edges in a weighted graph, the time-requirements of jobs to be scheduled, or the
sequence of molecular weights of a spectrographic dataset. Our investigations are of importance for problem input distributions
where the number
of distinct numerical values is small compared to the overall input size, and when the modeling of the problem allows rounding as a way
to get to fewer distinct values.

In classical complexity, this ``parameterization'' has been explored in dis\-tri\-bu\-tion-sensitive algorithmics \cite{SenG99}.
For example, while $\Omega(n \log n)$ is a lower bound on sorting $n$ values in the comparison model \cite{Knuth73},
a multiset of cardinality $n$ and $h$ distinct values can be sorted using $O(n \log h)$ comparisons \cite{MunroS76}.

It is perhaps surprising that this parameterization in the sense of Niedermeier's ``art of problem
parameterization'' \cite{Niedermeier06,Niedermeier10} has not been considered before in parameterized complexity, as it seems entirely
well-motivated.  While weighted combinatorial optimization problems have generally strong claims
to model realism, it is often the case that, e.g., the jobs to be scheduled may be of certain standard
sizes arising in a limited number of ways, or that the costs of the nodes in a network problem may
depend on the model and vendor of the device, of which there are a limited number of possibilities.
Many similar scenarios easily come to mind.  A bounded number of numbers may also arise naturally
and implicitly in parameterized problems when numbers are associated to other parameterized aspects
of a problem, such as alphabet size.

As an initial foray, we first show that a number of classic 
NP-hard problems about multisets of integers, when parameterized in this way, become 
\fpt.  The proofs are easy, and the knowledgeable reader might anticipate them almost as exercises today --- they use the relatively deep result that
{\sc Integer Linear Programming}, parameterized by the number of variables, is FPT.
Until recently, as noted in the 2006 monograph by Niedermeier \cite{Niedermeier06}, there were not so many 
interesting applications of this fundamental result (see \cite{AlonAWY98,FellowsLMRS08,FialaGK11,GrammNR03} for some exceptions).

At a deeper level of engagement with this parameterization, we describe some examples of how
number-of-numbers parameterized problems reduce to numerical problems about Mealy
machines, parameterized by the size of the description of the machine.  We show that
one basic problem about Mealy machines, parameterized in this way, is FPT, and that another is
$W[1]$-hard.

\section{Preliminaries}

\paragraph{Integer Linear Programming}
In the \ILPF problem (\sILPF), the input is an $m \times n$ matrix $\mathbf{A}$ of integers and an $m$-vector $\mathbf{b}$ of integers, the parameter is $n$, and the question is whether there exists an $n$-vector $\mathbf{x}$ of integers satisfying the $m$ inequalities $\mathbf{A} \mathbf{x} \le \mathbf{b}$.
\sILPF, parameterized by the number of variables, was shown to be \fpt by Lenstra~\cite{Lenstra83} and the running time has been improved by Kannan~\cite{Kannan87} and by Frank and Tardos~\cite{FrankT87}.

\paragraph{Multisets}
Let $A$ be a multiset. The \emph{cardinality} of $A$, denoted $|A|$, is the total number of elements in $A$, including repeated memberships. The \emph{variety} of $A$, denoted $||A||$, is the number of distinct elements in $A$. Element $a$ has \emph{multiplicity} $m$ in $A$ if it occurs $m$ times in $A$.
We denote the set of integers from $1$ to $n$ by $[n]\longversion{ = \set{1, \hdots, n}}$.

\paragraph{Graphs}
Let $G=(V,E)$ be a graph, $v\in V$ be a vertex of $G$, and $S\subseteq V$ be a subset of vertices of $G$. The subgraph of $G$ induced on $S$ is the graph $G[S] = (S,E\cap \set{uv: u,v\in S})$. The set $S$ is a \emph{clique} of $G$ if $G[S]$ is \emph{complete}, i.e. there is an edge between every two distinct vertices of $G[S]$. The set $S$ is an \emph{independent set} of $G$ if $G[S]$ is \emph{empty}, i.e. $G[S]$ has no edge. The \emph{neighborhood} of $v$ is the set of vertices incident to $v$ and denoted $N(v)$. The \emph{degree} of $v$ is $d(v) = |N(v)|$. We also define $N_S(v) = N(v)\cap S$ and $d_S(v)=|N_S(v)|$.

\paragraph{Words}
Let $\Sigma$ be an \emph{alphabet}. The elements of $\Sigma$ are called \emph{letters}, and a \emph{word} $x$ of length $n=|x|$ is a sequence of $n$ letters.  The symbol $\epsilon$ denotes the empty letter. We denote the concatenation of two words $x_1,x_2\in \Sigma^*$ by $x_1x_2$. The $i^{\text{th}}$ \emph{power} of a word $x$ is denoted $x^i$ or $(x)^i$ and represents the word $\underbrace{x x \dots x}_{i \text{ times}}$.

\paragraph{Parameterized Complexity}
We define the basic notions of Parameterized Complexity and refer to
other sources~\cite{DowneyF99,FlumG06,Niedermeier06} for an in-depth
treatment. A \emph{parameterized problem} is a set of pairs
$(I,k)$, the instances, where $I$ is the main part and $k$ is the
parameter. A parameterized problem is \emph{\fpt{}} if there
exist a computable function $f$ and an algorithm that solves any instance $(I,k)$ of size $n$ in time
$f(k)n^{O(1)}$. $\FPT$ denotes the
class of all \fpt parameterized decision problems.

Parameterized complexity offers a completeness theory that allows the accumulation of strong
theoretical evidence that some parameterized problems are not
\fpt. This theory is based on a hierarchy of
complexity classes 
\begin{align*}
\FPT \subseteq \W[1] \subseteq \W[2] \subseteq \W[3] \subseteq \cdots \subseteq \XP. 
\end{align*}
where all inclusions are believed to be strict. Each class $\W[i]$
contains all parameterized decision problems that can be reduced to a
canonical parameterized satisfiability problem $P_i$ under
\emph{\fptred{}s}.  These are many-to-one reductions where the parameter
for one problem maps into the parameter for the other. More
specifically, a parameterized problem $L$ reduces to a parameterized
problem $L'$ if there is a mapping $R$ from instances of~$L$ to
instances of $L'$ such that
\begin{enumerate}
\item $(I,k)$ is a \textsc{Yes}-instance of $L$ if and only if $(I',k')=R(I, k)$ is a \textsc{Yes}-instance
of~$L'$,
\item there is a computable function $g$ such that $k'\le g(k)$, and
\item there is a computable function $f$
such that $R$ can be computed in time $f(k) \cdot n^{O(1)}$, where
$n$ denotes the size of $(I,k)$.
\end{enumerate}
A parameterized
problem $L$ is then in $\W[i]$, for $i\in \mathbb{N}$, if it has a \fptred to the
problem of deciding whether a Boolean decision circuit (a decision
circuit is a circuit with exactly one output), with AND, OR, and NOT
gates, of constant depth such that on each path from an input to the
output, all but $i$ gates have a constant number of inputs,
parameterized by the number of ones in a satisfying assignment to the
inputs of the circuit~\cite{DowneyF99}.

A parameterized problem is in $\XP$ if there
exist computable functions $f$ and $g$ and an algorithm that solves any instance $(I,k)$ of size $n$ in time
$f(k)n^{g(k)}$.

\section{Subset Sum and Partition}

We start with two classic problems on multisets and show that they are \fpt, parameterized by the number of numbers.

\pbDef{\SubsetSum (\sSubsetSum)}
{A multiset $A$ of integers and an integer $s$.}
{$k = ||A||$, the number of distinct integers in $A$.}
{Is there a multiset $X\subseteq A$ such that $\sum_{a\in X} a = s$?}

\pbDef{\Partition (\sPartition)}
{A multiset $A$ of integers.}
{$k = ||A||$.}
{Is there a multiset $X\subseteq A$ such that $\sum_{a\in X} a = \sum_{b\in A\setminus X} b$?}

\noindent
The parameterizon of \textsc{Subset Sum} by $|X|$ is $W[1]$-hard \cite{FellowsK93}. This hardness also holds
for the parameterization of \textsc{Partition} by $|X|$ as an easy reduction from \textsc{Subset Sum} adds the integer $(\sum_{a \in A} a)-2s$ to $A$
if $s\le (\sum_{a \in A} a)/2$, and if $s> (\sum_{a \in A} a)/2$, the reduction looks instead for the complement set $A\setminus X$
that sums to $(\sum_{a \in A} a) -s$ and uses the previous construction.


Our \FPT\ results use a deep result of Lenstra, stating that \ILPF (\sILPF), parameterized by the number of variables, is FPT.
They are obtained by very natural formulations of the respective problems as integer programs.

\begin{theorem}\label{thm:SSfpt}
\sSubsetSum is \fpt.
\end{theorem}
\begin{proof}
Given an instance $(A,s)$ for \sSubsetSum, with $||A||=k$, we create an equivalent instance of \sILPF whose number of variables is upper bounded by a function of $k$.
Let $a_1, \hdots, a_k$ denote the distinct elements of $A$ and let $m_1, \hdots, m_k$ denote their respective multiplicities in $A$.
The \sILPF instance has the integer variables $x_1, \hdots, x_k$ and the following inequalities and equalities.
\begin{align*}
 x_i &\le m_i & \forall i \in [k]\\
 x_i &\ge 0 & \forall i \in [k]\\
 \sum_{i=1}^k x_i \cdot a_i &= s.
\end{align*}
For each $i\in [k]$, the variable $x_i$ represents the number of times $a_i$ occurs in $X$, the set summing to $s$ in a valid solution.
Using standard techniques in mathematical programming, these constraints can be transformed into the form $\mathbf{A} \mathbf{x} \le \mathbf{b}$.
\qed \end{proof}

\noindent
A very similar proof shows that \sPartition is \fpt. 

\begin{theorem}\label{thm:Pfpt}
\sPartition is \fpt.
\end{theorem}
\longversion{ %
  \begin{proof}
  Given an instance $A$ for \sPartition, with $||A||=k$, we create an equivalent instance of \sILPF whose number $n$ of variables is upper bounded by a function of $k$.\\
Let $a_1, \hdots, a_k$ denote the distinct elements of $A$ and let $m_1, \hdots, m_k$ denote their respective multiplicities in $A$.
The \sILPF instance has the integer variables $x_1, \hdots, x_k$ and the following inequalities and equalities.
\begin{align*}
 x_i &\le m_i & \forall i \in [k]\\
 x_i &\ge 0 & \forall i \in [k]\\
 \sum_{i=1}^k x_i \cdot a_i &= \sum_{a\in A} a / 2.
\end{align*}
For each $i\in [k]$, the variable $x_i$ represents the number of times $a_i$ occurs in $X$, such that $\sum_{a\in X} a = \sum_{b \in A\setminus X} b = \sum_{a\in A} a / 2$ in a valid solution.\\
Using standard techniques in mathematical programming, these constraints can be transformed such that they respect the form $\mathbf{A} \mathbf{x} \le \mathbf{b}$.

  \qed \end{proof}
}

\section{Other Classic Numerical Problems}

Using the \sILPF machinery, we show in this section that several other problems, which are often used in NP-hardness proofs, become \fpt when parameterized by the number of numbers.

\pbDef{\NumericalTDMatching (\sNumericalTDMatching)}
{Three multisets $A,B,C$ of $n$ integers each and an integer $s$.}
{$k = ||A\cup B\cup C||$.}
{Are there $n$ triples $S_1, \hdots, S_n$, each containing one element from each of $A,B,$ and $C$ such that for every $i\in [n]$, $\sum_{a\in S_i} a = s$?}

\begin{theorem}\label{thm:NTDMfpt}
\sNumericalTDMatching is \fpt.
\end{theorem}
\begin{proof}
Let $(A,B,C,s)$ be an instance for \sNumericalTDMatching, with $k_1=||A||$, $k_2=||B||$, $k_3=||C||$, and $k=||A\cup B\cup C||$.
Let $a_1, \hdots, a_{k_1}$ denote the distinct elements of $A$, $b_1, \hdots, b_{k_2}$ denote the distinct elements of $B$, and $c_1, \hdots, c_{k_3}$ denote the distinct elements of $C$. Also, let $m_{1,a}, \hdots, m_{k_1,a},\linebreak[1] m_{1,b}, \hdots, m_{k_2,b},\linebreak[1] m_{1,c}, \hdots, m_{k_3,c}$ denote their respective multiplicities in $A$, $B$, and $C$.
We create an instance of \sILPF with at most $k^3$ integer variables $x_{i,j,\ell}$, for $i\in [k_1], j\in [k_2], \ell\in [k_3]$:
\begin{align*}
x_{i,j,\ell} &= 0 &\text{for each }(i,j,\ell)\in [k_1] \times [k_2] \times [k_3]\\
 & &\text{ such that } a_i+b_j+c_\ell \not = s\\
\sum_{(j,\ell)\in ([k_2],[k_3])} x_{i,j,\ell} &= m_{i,a} & \forall i\in [k_1]\\
\sum_{(i,\ell)\in ([k_1],[k_3])} x_{i,j,\ell} &= m_{j,b} & \forall j\in [k_2]\\
\sum_{(i,j)\in ([k_1],[k_2])} x_{i,j,\ell} &= m_{\ell,c} & \forall \ell\in [k_3]
\end{align*}
A variable $x_{i,j,\ell}$ represents the number of times the elements $a_i\in A$, $b_j\in B$ and $c_\ell\in C$ are used together to form a triple summing to $s$. The first constraint makes sure that such a triple is formed only if it sums to $s$. The remaining equalities make sure that each element of $A\cup B\cup C$ appears in a triple. Thus $n$ such triples are formed, all summing to $s$ if the integer program is feasible.
\qed \end{proof}

\noindent
Note that the problem is also \fpt if parameterized by $||A\cup B||$ only: we face a \textsc{No}-instance if $||C|| > ||\set{a+b : a\in A, b\in B}||$.
A closely related, well known numerical problem, is the following.

\pbDef{\NumericalMatchingTargetSums (\sNumericalMatchingTargetSums)}
{Three multisets $A,B,S$ of $n$ integers each.}
{$k = ||A\cup B\cup S||$.}
{Are there $n$ triples $C_1, \hdots, C_n \in A \times B \times S$, such that the $A$-element and the $B$-element from each $C_i$ sum to its $S$-element?}

\begin{corollary}
\sNumericalMatchingTargetSums is \fpt.
\end{corollary}

\noindent
By the previous discussion, the natural parameterization by $|| A\cup B ||$ is also \fpt.
A straightforward adaptation of the proof of Theorem \ref{thm:NTDMfpt} shows that \ThreePartition is \fpt. 

\pbDef{\ThreePartition (\sThreePartition)}
{A multiset $A$ of $3 n$ integers.}
{$k = ||A||$.}
{Are there $n$ triples $S_1, \hdots, S_n \subseteq A$, all summing to the same number?}

\begin{theorem}\label{thm:TPfpt}
\sThreePartition is \fpt.
\end{theorem}
\longversion{ %
  \begin{proof}
  Let $A$ be an instance for \sThreePartition, with $||A||=k$ and $|A|=3n$. Let $s = \sum_{a\in A} a / n$.
Let $a_1, \hdots, a_{k}$ denote the distinct elements of $A$ and let $m_{1}, \hdots, m_{k}$ denote their multiplicities in $A$.
We create an instance of \sILPF with at most $k^3$ integer variables $x_{i,j,\ell}$, for $i,j,\ell \in [k]$:
\begin{align*}
x_{i,j,\ell} &= 0 &\text{for each } i,j,\ell\in [k]\\
 & &\text{ such that } a_i+a_j+a_\ell \not = s\\
\sum_{\substack{j,\ell \in [k]\\j,\ell\not = i}} (x_{i,j,\ell} + x_{j,i,\ell} + x_{j,\ell ,i})\\
+ 2\cdot \sum_{\substack{j \in [k]\\j\not = i}} (x_{i,i,j} + x_{i,j,i} + x_{j,i,i})\\
+ 3\cdot x_{i,i,i} &= m_{i} & \forall i\in [k]\\
\end{align*}
A variable $x_{i,j,\ell}$ represents the number of times the elements $a_i,a_j$ and $a_\ell$ are used together to form a triple summing to $s$. The first constraint makes sure that such a triple is formed only if it sums to $s$. The second set of equalities make sure that each element of $A$ appears in a triple. Thus $n$ such triples are formed, all summing to $s$ if the integer program is feasible.

  \qed \end{proof}
}

\section{Mealy Machines}

In this section, we explore how far we can generalize the rather simple \FPT\ results of the previous two sections.
To this end, we investigate the parameterized complexity of two problems about Mealy Machines.
Both problems can be viewed as parameterized problems implicitly
parameterized by the number of numbers, because in each case the size of the alphabet is part of the
parameterization, and each letter of the alphabet is associated with a census requirement. The richer
structure of these problems means that a simple appeal to integer linear programming no longer suffices:
one turns out to be \FPT, and the other \W[1]-hard. In Section~\ref{sec:applications}, we show that other problems parameterized
by the number of numbers reduce to these two seemingly general problems of this kind.

Mealy machines \cite{Mealy55} are finite-state transducers, generating an output based on their current state and input.
They have important applications in cryptanalysis \cite{Bard09}, computational linguistics \cite{RocheS97}, and control
and system theory \cite{Sontag98}.
A \emph{\deterministic Mealy machine} is a dual-alphabet state transition system given by a $5$-tuple $M=(S,s_0,\Gamma, \linebreak[1] \Sigma,T)$:
\begin{itemize}
 \item a finite set of states $S$,
 \item a start state $s_0 \in S$,
 \item a finite set $\Gamma$, called the input alphabet,
 \item a finite set $\Sigma$, called the output alphabet, and
 \item a transition function $T: S \times \Gamma \rightarrow S \times \Sigma$ mapping pairs of a state and an input letter to the corresponding next state and output letter.
\end{itemize}
The alphabets $\Gamma$ and $\Sigma$ may contain the empty letter $\epsilon$, as in \cite{Savage98}. This eases some of the description, but all our results also hold if we restrict $\epsilon \notin \Gamma \cup \Sigma$.

In a \emph{\nondet Mealy machine}, the only difference is that the transition function is defined $T: S \times \Gamma \rightarrow \mathcal{P}(S \times \Sigma)$ as for a given state and input letter, there may be more than one possibility for the next state and output letter.  (Here $\mathcal{P}(X)$ denotes the powerset of a set $X$.) 

A \emph{census requirement} $c: \Sigma \setminus \set{\epsilon} \rightarrow \mathbb{N}$ is a function assigning a non-negative integer to each letter of the output alphabet (except $\epsilon$). It is used to constrain how many times each letter should appear in the output of a\longversion{ Mealy} machine. A word $x \in \Sigma^*$ \emph{meets} the census requirement if every letter $b\in \Sigma \setminus \{ \epsilon \}$ appears exactly $c(b)$ times in $x$.

The notion of census requirement is related to Parikh images \cite{Parikh66}.
Let $\Sigma \setminus \{\epsilon\} = \{b_1, \ldots, b_\sigma\}$. For $x\in \Sigma^*$, the \emph{Parikh image} is
$\Psi(x) = (c(b_1), \ldots, c(b_\sigma))$, where $c$ is the census requirement such that $x$ meets $c$.
The \emph{Parikh image of a language} $L$ is $\Psi(L) = \{\Psi(x) : x\in L\}$.
Parikh's theorem \cite{Parikh66} states that the Parikh image of a context-free language is semilinear, i.e., 
that for every context-free language there is a regular language with the same Parikh image.

Our first problem about Mealy machines asks whether there exists an input word and a computation of the Mealy machine such that the output word meets the census requirement.

\pbDef{\ExistsWordMM (\sExistsWordMM)}
{A \nondet Mealy machine $M=(S,s_0,\Gamma,\Sigma,T)$, and
a census requirement $c: \Sigma \setminus \set{\epsilon} \rightarrow \mathbb{N}$.}
{$|S|+|\Gamma|+|\Sigma|$.}
{Does there exist a word $x \in \Gamma^*$ for which a computation of $M$ on input $x$ generates an output $y$ that meets $c$?}

\noindent
Our proof that \sExistsWordMM is \fpt is inspired by the 
proof from \cite{FellowsLMMRS09} showing that \textsc{Bandwidth} is 
\fpt when parameterized by the maximum number of 
leaves in a spanning tree of the input graph. We need the following 
definition and lemma from \cite{FellowsLMMRS09}.

In a digraph $D$, two directed walks $\Delta$ and $\Delta '$ from 
a vertex $s$ to a vertex $t$ are \emph{arc-equivalent}, if for every 
arc $a$ of $D$, $\Delta$ and $\Delta '$ pass through $a$ the 
same number of times.

\begin{lemma}[\cite{FellowsLMMRS09}]\label{lem:walks}
Any directed walk $\Delta$ through a finite digraph $D$ on $n$ vertices 
from a vertex $s$ to a vertex $t$ of $D$ is arc-equivalent to a directed 
walk $\Delta '$ from $s$ to $t$, where $\Delta '$ has the form:
\begin{itemize}
\item[(1)] $\Delta '$ consists of an underlying directed walk $\rho$ 
from $s$ to $t$ of length at most $n^2$,
\item[(2)]  together with some number of \emph{short loops}, where each such short loop 
$l$ begins and ends at a vertex of $\rho$, and has length at most $n$.
\end{itemize}
\end{lemma}

\noindent
The algorithm will first subdivide state transitions in order to make the underlying directed graph simple.
As suggested by Lemma \ref{lem:walks}, the algorithm goes over all possible choices for selecting an underlying directed walk $\rho$ starting from $s_0$.
For every short loop starting and ending at a vertex from $\rho$, the algorithm associates an integer variable
representing the number of times this short loop is executed while moving along $\rho$. Again by \ILPF, it can be
checked whether there is a set of integers, representing the number of executions of the short loops, such that the number
of times each output letter is written is compatible with the census requirement.

\begin{theorem}\label{thm:mealyfpt}
 \sExistsWordMM is \fpt.
\end{theorem}
\begin{proof}
Let $(M'=(S',s_0',\Gamma',\Sigma',T'),c)$ be an instance for \sExistsWordMM with $k=|S'|+|\Gamma'|+|\Sigma'|$. As $M'$ might have multiple transitions from one state to another, we first subdivide each transition in order to obtain a simple digraph underlying the Mealy machine (so we can use Lemma \ref{lem:walks}): create a new \nondet Mealy machine $M=(S,s_0,\Gamma,\Sigma,T)$ such that, initially, $S=S'$, $s_0=s_0'$, $\Gamma = \Gamma' \cup \set{\epsilon}$, and $\Sigma = \Sigma '\cup \set{\epsilon}$; for each transition $t$ of $T'$ from a couple $(s_i, \letter{i})$ to a couple $(s_o,\letter{o})$, add a new state $s_{t}$ to $S$ and add the transition from $(s_i, \letter{i})$ to $(s_{t}, \letter{o})$ and the transition from $(s_{t}, \epsilon)$ to $(s_o, \epsilon)$ to $T$. Clearly, there is at most one transition between every two states in $M$.

Our algorithm goes over all transition walks in $M$ of length at most $|S|^2$ that start from $s_0$. There are at most $|S|^{(|S|^2)}$ such transition walks and each such transition walk has at most $|S|^{|S|}$ short loops, as they have length at most $|S|$ by Lemma \ref{lem:walks}. Let $P=(s_0,s_1,\hdots, s_{|P|})$ be such a transition walk and $L=(\ell_1, \ell_2, \hdots, \ell_{|L|})$ be its short loops. It remains to check whether there exists a set of integers $X=\set{x_1, x_2, \hdots, x_{|L|}}$ such that a word output by a computation of $M$ moving from $s_0$ to $s_{|P|}$ along the walk $P$, and executing $x_i$ times each short loop $\ell_i$, $1\le i\le |L|,$ meets the census requirement. Note that if one such word meets the census requirement, then all such words meet the census requirement, as it does not matter in which order the short loops are executed. We verify whether such a set $X$ exists by \sILPF.

Let $\Sigma \setminus \set{\epsilon} = \set{\letter{\ell,1},\letter{\ell,2}, \hdots, \letter{\ell,\sigma}}$. Define $m(i,j)$, for $1 \le i \le |L|$, $1\le j\le \sigma$, to denote the number of times that $M$ writes the letter $\letter{\ell,j}$ when it executes the loop $\ell_i$ once. Define $m(j)$, for $1\le j\le \sigma$, to be the number of times that $M$ writes the letter $\letter{\ell,j}$ when it transitions from $s_0$ to $s_{|P|}$ along the walk $P$. Then, we only need to verify that there exist integers $x_1, x_2, \hdots, x_{|L|}$ such that
\begin{align*}
 m(j) + \sum_{i=0}^{|L|} x_i \cdot m(i,j) &= c(\letter{\ell,j}), & \forall j \in [\sigma].
\end{align*}
By construction, $|S|\le |S'|+|T'| \le |S'| + |S'|^2 \cdot |\Gamma'|\cdot |\Sigma'| \le k+k^4$.
As the number of integer variables of this program is at most $|L| \le |S|^{|S|} \le (k+k^4)^{k+k^4}$, and the number of transition walks that the algorithm considers is at most $|S|^{(|S|^2)} \le (k+k^4)^{k^2+2k^5+k^8}$, \sExistsWordMM is \fpt.
\qed \end{proof}

\noindent
We note that the proof in \cite{FellowsLMMRS09} concerned a special case 
of a \deterministic Mealy machine where the input and output alphabet 
are the same, and all transitions that read a letter 
$\letter{\ell}$ also write $\letter{\ell}$.

\smallskip

In our second Mealy machine problem, the question is whether, for a given input word, there is a computation of the Mealy machine which outputs a word that meets the census requirement.

\pbDef{\GivenWordMM (\sGivenWordMM)}
{A \nondet Mealy machine $M=(S,s_0,\Gamma,\Sigma,T)$,
a word $x\in \Gamma^*$, and
a census requirement $c: \Sigma \setminus \{\epsilon\} \rightarrow \mathbb{N}$.}
{$|S|+|\Gamma|+|\Sigma|$}
{Is there a computation of $M$ on input $x$ generating an output $y$ that meets $c$?}

\noindent
\shortversion{Membership in $\XP$ is easily shown by dynamic programming.}%
\longversion{By dynamic programming we show that two restrictions of this problem are in $\XP$.
In the first one, the census requirement is encoded in unary.
This restriction of the problem seems lenient, especially when one is actually interested in finding the output word, as the
census function acts then as a placeholder for the produced word.}

\begin{theorem}\label{thm:dynprog}
\sGivenWordMM is in $\XP$ if $c$ is encoded in unary.
\end{theorem}
\longversion{ %
  \begin{proof}
  Let $|\Sigma \setminus \set{\epsilon}| = \sigma$ and $\Sigma \setminus \set{\epsilon} = \set{b_1, \hdots, b_{\sigma}}$. Our dynamic programming algorithm computes the entries of a boolean table $A$. The table $A$ has an entry $A[s, c_1, \linebreak[1] \hdots, \linebreak[1] c_{\sigma}, i, p]$ for each state $s\in S$, each $c_j \in \set{0, \hdots, c(b_j)}$, $j\in [\sigma]$, each index $i \in \{0, \ldots, |x|\}$, and each integer $p\in P=\{0, \ldots, |S|-1\}$. The entry $A[s, c_1, \hdots, c_{\sigma}, i, p]$ is set to \texttt{true} if there exists a computation of $M$ reading the first $i$ letters of $x$, outputting a word $y$ in which the letter $b_j$ occurs $c_j$ times, for each $j\in [\sigma]$, followed by $p$ transitions that read $\epsilon$ and write $\epsilon$, and ending up in state $s$, and to \texttt{false} otherwise.\\
Set $A[s, c_1, \hdots, c_{\sigma}, 0, 0]$ to \texttt{true} if $s=s_0$ and $c_1 = \hdots = c_{\sigma} = 0$, and to \texttt{false} otherwise. Compute the values of the table by increasing values of $\sum_{i=1}^\sigma c_i$, index $i$, propagation integer $p$, and state number $s$:
\begin{alignat*}{2}
 &A[s, c_1, \hdots, c_{\sigma}, i, p] =&\\ &\quad \bigvee_{\genfrac{}{}{0pt}{}{s'\in S,b_j \in \Sigma\setminus \set{\epsilon}, p'\in P :}{(s,b_j) \in T(s',x[i])}} && A[s', c_1, \hdots, c_{j-1},c_j-1,c_{j+1}, \hdots, c_{\sigma}, i-1,p']\\
                                  &\vee \;\:\:\:\, \bigvee_{\genfrac{}{}{0pt}{}{s'\in S, p'\in P :}{(s,\epsilon) \in T(s',x[i])}} && A[s', c_1, \hdots, c_{\sigma}, i-1, p']\\
                                  &\vee \:\! \bigvee_{\genfrac{}{}{0pt}{}{s'\in S,b_j \in \Sigma\setminus \set{\epsilon}, p'\in P :}{(s,b_j)\in T(s',\epsilon)}} && A[s', c_1, \hdots, c_{j-1},c_j-1,c_{j+1}, \hdots, c_{\sigma}, i,p']\\
                                  &\vee \:\, \bigvee_{s'\in S :(s,\epsilon)\in T(s',\epsilon)} && A[s', c_1, \hdots, c_{\sigma}, i, p-1]\\
\end{alignat*}
Finally, there exists an $x$-computation of $M$ generating a word $y$ that meets the census requirement if and only if $\bigvee_{s \in S, p\in P} A[s,c(b_1), \hdots, c(b_{\sigma}), |x|, p]$ is \texttt{true}. Denote by $n$ is the length of the description of an input instance.
The table has $|S|\cdot |x|\cdot |P|\cdot \Pi_{j=1}^{\sigma} c(b_{j}) \le |S|^2 \cdot n^{\sigma+1}$ entries, and each entry can be computed in time $O(|S|^2 \cdot \sigma)$. The running time of the algorithm is thus upper bounded by $O(n^{\sigma+1}\cdot k^5)$, where $k$ is the parameter.

  \qed \end{proof}
}

\noindent
For the version where $c$ is encoded in binary, a restriction on the input alphabet gives an $\XP$ algorithm as well.

\begin{corollary}\label{thm:dynprog2}
\sGivenWordMM is in $\XP$ if $\epsilon \notin \Gamma$.
\end{corollary}
\begin{proof}
 If $\sum_{b\in \Sigma \setminus \{\epsilon\}} c(b) > |x|$, then return \texttt{false}, as $M$ cannot output more than $|x|$ letters.
 Otherwise, run the algorithm described in the proof of Theorem \ref{thm:dynprog}.
 Its running time is $O(k^5\cdot |x| \cdot \Pi_{j=1}^{\sigma} c(b_{j})) = O(n^{\sigma+1} \cdot k^5)$.
\qed \end{proof}

\noindent
Note that the \XP-results also hold if the parameter is only $|\Sigma|$.

\medskip
\shortversion{To show $W[1]$-hardness, we reduce from \sMCC, which is $W[1]$-hard \cite{Pietrzak03,FellowsHRV09}.}%
\longversion{To show that \sGivenWordMM is $W[1]$-hard, we reduce from the \MCC problem, which is $W[1]$-hard \cite{Pietrzak03,FellowsHRV09}.}

\pbDef{\MCC (\sMCC)}
{An integer $k$ and a connected undirected graph $G=(V(1)\cup V(2) \hdots \cup V(k),E)$ such that for every $i\in [k]$, the vertices of $V(i)$ induce an independent set in $G$.}
{$k$.}
{Is there a clique of size $k$ in $G$?}

\noindent
Clearly, a solution to this problem has one vertex from each color.

Our \fptred encodes $G$ in the input word $x$ of the Mealy machine $M$, and the description of $M$ depends only on $k$.
The Mealy machine is divided into $k$ parts, one for each color class $V(i)$, with $1\le i\le k$. Its $i^{\text{th}}$ part is responsible
for selecting a vertex $v_i$ from $V(i)$ and edges $v_i v_j$ for every $v_j \in V(j)$, with $1\le j \neq i \le k$.
All consistency issues and communication is done via the census requirement.
Within part $i$, we need to make sure that the selected edges are all incident to the selected vertex $v_i$. This is achieved by making
$M$ output $p$ times each letter $\letter{\mathbf{l},i,j}$, with $1\le j\neq i \le k$, if it selects the $p^{\text{th}}$ vertex in $V(i)$.
The census requirement for $\letter{\mathbf{l},i,j}$ is $|V(i)|+1$, meaning that $\letter{\mathbf{l},i,j}$ needs to be output $|V(i)|+1-p$ times later.
To select an edge $v_i v_j$, the machine $M$ will be constrained to select this edge among the edges incident to $v_i$. To achieve this,
edges from $V(i)$ to $V(j)$ appear in $x$ grouped by the vertex from $V(i)$ on which they are incident. After each group of edges
incident on one vertex from $V(i)$, there is a special state where $\letter{\mathbf{l},i,j}$ is output if and only if the edge towards $V(j)$ has already been selected.
As $\letter{\mathbf{l},i,j}$ needs to be output exactly $|V(i)|+1-p$ times, we force in this way that an edge is selected which is incident on $v_i$.
This enforces that all edges selected in the $i^{\text{th}}$ part are incident on the same vertex. It remains to make sure that distinct parts
$i$ and $j$ select the same edge between $V(i)$ and $V(j)$. This is again achieved by a census requirement where a part of the census of letter
$\letter{\mathbf{l},\bar{\mathbf{e}},i,j}$ is output in the $i^{\text{th}}$ part and the remaining part in the $j^{\text{th}}$ part of $M$.

\begin{theorem}
\sGivenWordMM is $W[1]$-hard.
\end{theorem}
\begin{proof}
Let $(k,G=(V(1)\cup V(2) \hdots \cup V(k),E))$ be an instance of \sMCC. Suppose $V(i)=\set{v_{i,1}, v_{i,2}, \hdots, v_{i,|V(i)|}}$ is the vertex set of color $i$, for each color class $i\in [k]$, $E=\set{e_1, e_2, \hdots, e_{|E|}}$, and $E(i,j) = \set{e(i,j,1), e(i,j,2), \linebreak[1] \hdots, e(i,j,|E(i,j)|)}$ is the subset of edges with one vertex in color class $i$ and the other in color class $j$, for $i,j\in [k]$. Moreover, suppose $E(i,j)$ follows the same order as $E$, that is if $e_p = e(i,j,p')$,  $e_q=e(i,j,q')$, and $p \le q$, then $p'\le q'$. For a vertex $v_{i,p}$ and two integers $j\in [k]\setminus \set{i}$ and $q\in [d_{V(j)}(v_{i,p})+1]$, we define $\gap(v_{i,p},j,q) = t-s$, where $e(i,j,t)$ is the $q^{\text{th}}$  edge in $E(i,j)$ incident to $v_{i,p}$ (respectively, $t=|E(i,j)|$ if $q = d_{V(j)}(v_{i,p})+1$) and $e(i,j,s)$ is the $(q-1)^{\text{th}}$ edge in $E(i,j)$ incident to $v_{i,p}$ (respectively, $s=0$ if $q=1$).

We construct an instance $(M=(S,s_0,\Gamma,\Sigma,T), x,c)$ for \sGivenWordMM\longversion{ as follows}. $M$'s input alphabet, $\Gamma$, is $\set{\letter{i}, \letter{i,j}, \letter{\bar{\mathbf{e}},i,j}, \letter{\mathbf{e},i,j} : i,j\in [k], i\not = j}$. $M$'s output alphabet, $\Sigma$, is $\set{\epsilon} \cup \set{\letter{\mathbf{l},i,j}, \letter{\mathbf{l},\bar{\mathbf{e}},i,j} : i,j\in [k], \linebreak[1] i\not = j}$. The word $x$ is defined

\begin{align*}
x  &:= x_1 x_2 \hdots x_k\\
x_i &:= x_{i,0} x_{i,1} \hdots x_{i,i-1} x_{i,i+1} x_{i,i+2} \hdots x_{i,k} \letter{i} & \forall i\in [k]\\
x_{i,0} &:= (\letter{i,1} \letter{i,2} \hdots \letter{i,i-1} \letter{i,i+1} \letter{i,i+2} \hdots \letter{i,k})^{|V(i)|} & \forall i\in [k]\\
x_{i,j} &:= \letter{i,j} x_{i,j,1} \letter{i,j} x_{i,j,2}  \hdots \letter{i,j} x_{i,j,|V(i)|} \letter{i,j} &\shortversion{\hspace{-26pt}}\longversion{\hspace{-34.26pt}} \forall i,j\in[k], i \not = j\\
x_{i,j,p} &:= \letter{\bar{\mathbf{e}},i,j}^{\gap(v_{i,p},j,1)} \letter{\mathbf{e},i,j} \letter{\bar{\mathbf{e}},i,j}^{\gap(v_{i,p},j,2)} \letter{\mathbf{e},i,j}\\
        & \; \hdots \letter{\bar{\mathbf{e}},i,j}^{\gap(v_{i,p},j,d_{V(j)}(v_{i,p}))} \letter{\mathbf{e},i,j} \letter{\bar{\mathbf{e}},i,j}^{\gap(v_{i,p},j,d_{V(j)}(v_{i,p})+1)}.
\end{align*}
The census requirement $c$ is, for every $i,j\in [k], i\not = j$,
\begin{align*}
 c(\letter{\mathbf{l},i,j}) & := |V(i)|+1\\
 c(\letter{\mathbf{l},\bar{\mathbf{e}},i,j}) & := |E(i,j)|.
\end{align*}
On reading a subword $x_i$, the Mealy machine will select a vertex $v_{i,p}$ in $V(i)$ and one edge incident to $v_{i,p}$
for each color class $j \in [k]\setminus \{i\}$. The vertex $v_{i,p}$ is selected in the subword $x_{i,0}$ of $x_i$. Next, for each
$j \in [k]\setminus \{i\}$, a vertex in $V(i)$ and a vertex in $V(j)$ are selected in the subword $x_{i,j}$. The census
requirement for $\letter{\mathbf{l},i,j}$ makes sure that the vertex from $V(i)$ is $v_{i,p}$. The subword $x_{i,j,p}$
ensures that $v_{i,p}$ and the vertex that is selected from $V(j)$ are joined by an edge. Finally, the census requirement
for $\letter{\mathbf{l},\bar{\mathbf{e}},i,j}$ is responsible for the inter-partition communication and makes sure that the edge
selected in $x_{i,j}$ is equal to the edge selected in $x_{j,i}$.

The Mealy machine $M$ consists of $k$ parts. The $i^{\text{th}}$ part of $M$ is depicted in Fig. \ref{fig:automaton}. Its initial state is $s_{v,1}$. There is a transition from the last state of each part, $s_{e,i,k}^{(4)}$, to the first state of the following part, $s_{v,i+1}$ (from the $k^{\text{th}}$ part, there is a transition to a final state): it reads the letter $\letter{i}$ and writes the letter $\epsilon$. We set $\letter{\mathbf{l} ', \bar{\mathbf{e}},i,j} = \letter{\mathbf{l}, \bar{\mathbf{e}},j,i}$ for all $i \not = j \in [k]$.
Note that, in the description of $M$, the letter $\letter{\mathbf{l}, i,j}$ can only be output on reading $\letter{i,j}$, and $\letter{\mathbf{l}, \bar{\mathbf{e}},i,j}$ can only be output on reading $\letter{\bar{\mathbf{e}},i,j}$ or $\letter{\bar{\mathbf{e}},j,i}$.

\begin{figure}[t!]
\begin{center}
\longversion{\includegraphics[scale=0.92]{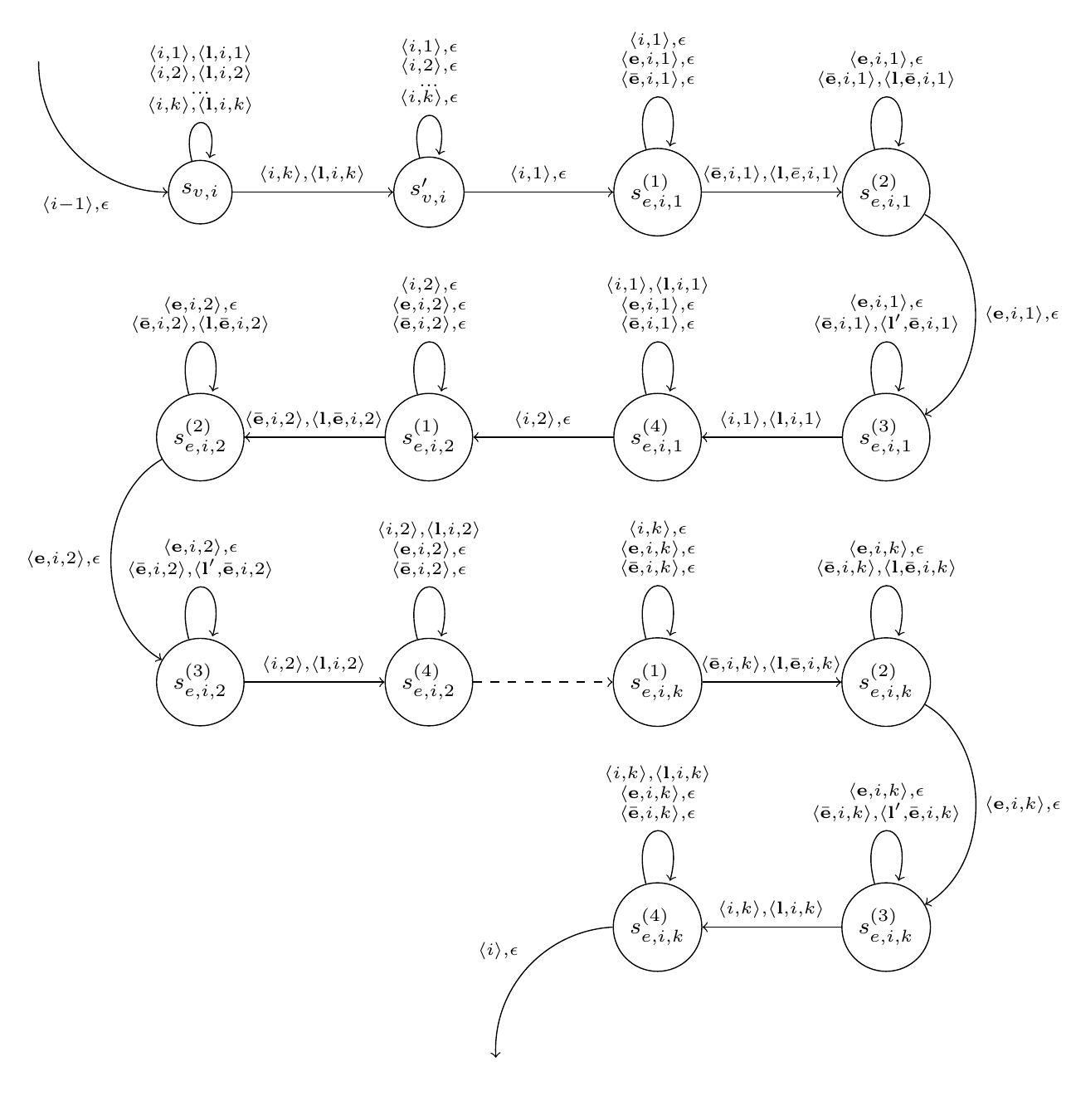}}
\caption{\label{fig:automaton}The $i^{\text{th}}$ part of the Mealy machine $M$.
It does not have the states $s_{e,i,i}^{(1)}, \protect\linebreak[1] s_{e,i,i}^{(2)}, \protect\linebreak[1] s_{e,i,i}^{(3)}, $ and $s_{e,i,i}^{(4)}$; there is instead a transition from $s_{e,i,i-1}^{(4)}$ to $s_{e,i,i+1}^{(1)}$ reading $\letter{i-1}$ and writing $\epsilon$, and there is a transition from $s_{e,k,k-1}^{(4)}$ to the last state reading $\letter{k}$ and writing $\epsilon$. There are no transitions starting at the last state. (Drawing all this would have cluttered the figure too much.)}
\end{center}
\end{figure}

\smallskip

Let us first verify that the parameter for \sGivenWordMM is a function of $k$, and that there exists a function $f$ such that the size of the instance for \sGivenWordMM is $f(k) \cdot n^{O(1)}$, where $n$ is the number of vertices of $G$.
We have $|\Gamma| = k+3\cdot k\cdot (k-1)$, $|\Sigma| = 1+2\cdot k\cdot (k-1)$, and $|S| = 1+k\cdot (2+4\cdot (k-1))$. The parameter of \sGivenWordMM is thus bounded by a function of $k$.
The length of $x$ is $O(k^2 \cdot n^3)$.
Now, we show that $(M,x,c)$ is a \textsc{Yes}-instance for \sGivenWordMM if and only if $(G,k)$ is a \textsc{Yes}-instance for \sMCC.

\medskip

\noindent
First, suppose $(M=(S,s_0,\Gamma,\Sigma,T), x,c)$ is a \textsc{Yes}-instance for \sGivenWordMM.

We say that $M$ \emph{selects} a vertex $v_{i,p}$ if it makes a transition from state $s_{v,i}$ to state $s'_{v,i}$ reading $\letter{i,k}$ (respectively $\letter{i,k-1}$ if $i=k$) for the $p^{\text{th}}$ time. In other words, in the $i^{\text{th}}$ part of $M$, it reads $p\cdot (k-1)-1$ letters of $x_{i,0}$, staying in state $s_{v,i}$ and outputs the letter $\letter{\mathbf{l},i,r}$ for each letter $\letter{i,r}$ it reads; then it transitions to state $s'_{v,i}$ on reading $\letter{i,k}$ (respectively $\letter{k,k-1}$ if $i=k$) and outputs $\letter{\mathbf{l},i,k}$ (respectively $\letter{\mathbf{l},k,k-1}$); in the state $s'_{v,i}$ it outputs the empty letter for each letter $\letter{i,r}$ it reads.

We say that $M$ \emph{selects} an edge $e(i,j,q)$ if it makes a transition from state $s_{e,i,j}^{(2)}$ to state $s_{e,i,j}^{(3)}$ after having read the letter $\letter{\bar{\mathbf{e}},i,j}$ of $x_{i,j,p}$ exactly $q$ times, where $v_{i,p}$ is the vertex of color $i$ that $e(i,j,q)$ is incident on. In other words, in the $i^{\text{th}}$ part of $M$, it transitions from the state $s_{e,i,j}^{(1)}$ to the state $s_{e,i,j}^{(2)}$ on reading the first letter of $x_{i,j,p}$ (if it did this transition any later, the census requirement of $\letter{\mathbf{l},\bar{\mathbf{e}},i,j}$ could not be met\longversion{, as shown in the proof of Claim \ref{cl:match-check} below}); then it stays in the state $s_{e,i,j}^{(2)}$ until it has read $q$ times the letter $\letter{\bar{\mathbf{e}},i,j}$ of $x_{i,j,p}$; then it transitions to the state $s_{e,i,j}^{(3)}$ on reading $\letter{\mathbf{e},i,j}$; it stays in this state and outputs $\letter{\mathbf{l} ',\bar{\mathbf{e}},i,j}$ for each letter $\letter{\bar{\mathbf{e}},i,j}$ it reads until transitioning to the state $s_{e,i,j}^{(4)}$ on reading the letter following $x_{i,j,p}$.

The following claims ensure that the edge-selection and the vertex-selection are compatible, i.e., that exactly one edge is selected from color $i$ to color $j$, and that this edge is incident on the selected vertex of color $i$.

\begin{boldclaim}\label{cl:coherence}
 Let $i$ be a color and let $v_{i,p}$ be the vertex selected in the $i^{\text{th}}$ part of $M$. In its $i^{\text{th}}$ part, $M$ selects one edge incident to $v_{i,p}$ and to a vertex of color $j$, for each $j\in [k]\setminus \set{i}$.
\end{boldclaim}
\begin{proof}
After $M$ has selected $v_{i,p}$, it has output $p$ times each of the letters $\letter{\mathbf{l},i,1}, \linebreak[1] \letter{\mathbf{l},i,2}, \linebreak[1] \hdots, \linebreak[1] \letter{\mathbf{l},i,i-1}, \linebreak[1] \letter{\mathbf{l},i,i+1}, \linebreak[1] \letter{\mathbf{l},i,i+2}, \linebreak[1] \hdots, \linebreak[1] \letter{\mathbf{l},i,k}$. For each $j \in [k] \setminus \set{i}$, the only other transitions that output $\letter{\mathbf{l},i,j}$ are the transition from $s_{e,i,j}^{(3)}$ to $s_{e,i,j}^{(4)}$ and a transition that loops on $s_{e,i,j}^{(4)}$. To meet the census requirement of $|V(i)|+1$ for $\letter{\mathbf{l},i,j}$, $M$ selects an edge while reading $x_{i,j,p}$. This edge is incident on $v_{i,p}$ by construction.
\qed \end{proof}

\noindent
The following claim makes sure that the edge selected from color $i$ to color $j$ is the same as the edge selected from color $j$ to color $i$.

\begin{boldclaim}\label{cl:match-check}
 Suppose $M$ selects the edge $e(i,j,q)$ in its $i^{\text{th}}$ part. Then, $M$ selects the edge $e(j,i,q)$ in its $j^{\text{th}}$ part.
\end{boldclaim}
\longversion{\begin{proof}
Before $M$ selects $e(i,j,q)$, it has output $q' \le q$ times the letter $\letter{\mathbf{l},\bar{\mathbf{e}},i,j}$. On selecting $e(i,j,q)$ it transitions to the state $s_{e,i,j}^{(3)}$, and after the selection it outputs $\letter{\mathbf{l} ', \bar{e}, i, j}$ for every letter $\letter{\bar{\mathbf{e}},i,j}$ of $x_{i,j,p}$ it reads. As it reads \begin{equation*}
(\sum_{r=1}^{d_{V(j)}(v_{i,p})+1} \gap(v_{i,p},j,r))-q = |E(i,j)|-q
\end{equation*}
times the letter $\letter{\bar{\mathbf{e}},i,j}$ of $x_{i,j,p}$ after it has selected $e(i,j,q)$, the Mealy machine outputs $|E(i,j)|-q$ times the letter $\letter{\mathbf{l} ', \bar{e}, i, j}$ in its $i^{\text{th}}$ part.

The only other transition where it outputs $\letter{\mathbf{l}, \bar{e}, i,j} = \letter{\mathbf{l} ', \bar{e} ,j,i}$ is the transition in the $j^{\text{th}}$ part of $M$ looping on $s_{e,j,i}^{(3)}$ that reads $\letter{\bar{\mathbf{e}},j,i}$ and outputs $\letter{\mathbf{l} ',\bar{\mathbf{e}},j,i}$. To meet the census requirement for $\letter{\mathbf{l}, \bar{e},i,j}$, this transition must be used exactly $|E(i,j)|-q'$ times.

The only other transitions where it outputs $\letter{\mathbf{l} ', \bar{e},i,j} = \letter{\mathbf{l}, \bar{e}, j, i}$ are two transitions in the $j^{\text{th}}$ part of $M$: the transition from $s_{e,j,i}^{(1)}$ to $s_{e,j,i}^{(2)}$ and the transition looping on $s_{e,j,i}^{(2)}$, both reading $\letter{\bar{\mathbf{e}},j,i}$ and writing $\letter{\mathbf{l}, \bar{e},j,i}$. These transitions can be used at most $q'$ times as the transition of the previous paragraph is used $|E(i,j)|-q'$ times. These transitions have to be used at least $q$ times to meet the census requirement for $\letter{\mathbf{l} ',\bar{\mathbf{e}},i,j}$. Thus, these transitions are used exactly $q$ times and $q=q'$.

Finally, the transition from $s_{e,j,i}^{(2)}$ to $s_{e,j,i}^{(3)}$ happens after having read $q$ times the letter $\letter{\bar{\mathbf{e}},j,i}$ of some vertex $x_{j,i,p'}, p'\in [|V(j)|]$, which means that $M$ selects the edge $e(j,i,q)$ in its $j^{\text{th}}$ part.
\qed \end{proof}}

\noindent
By Claims \ref{cl:coherence} and \ref{cl:match-check}, the $k$ vertices that are selected by $M$ form a multicolored clique. Thus, $(k,G=(V(1)\cup V(2) \hdots \cup V(k),E))$ is a \textsc{Yes}-instance for \sMCC.

\bigskip

\noindent
Now, suppose that $(k,G=(V(1)\cup V(2) \hdots \cup V(k),E))$ is a \textsc{Yes}-instance for \sMCC.

Let $\set{v_{1,p_1},v_{2,p_2}, \hdots, v_{k,p_k}}$ be a multicolored clique in $G$. We will construct a word $y$ meeting $c$ such that a computation of $M$ on input $x$ generates $y$. For two adjacent vertices $v_{i,p_i}$ and $v_{j,p_j}$, define $\edge(v_{i,p_i},v_{j,p_j}) = t$ such that $e(i,j,t) = v_{i,p_i} v_{j,p_j}$. The word $y$ is $y_1 y_2 \hdots y_k$, where $y_i$, for $i\in [k]$, is
\begin{align*}
 &(\letter{\mathbf{l},i,1} \letter{\mathbf{l},i,2} \hdots \letter{\mathbf{l},i,i-1} \letter{\mathbf{l},i,i+1} \letter{\mathbf{l},i,i+2} \hdots \letter{\mathbf{l},i,k})^{p_i}\\
 &y_{i,1} y_{i,2} \hdots y_{i,i-1} y_{i,i+1} y_{i,i+2} \hdots y_{i,k} 
\end{align*}
and $y_{i,j}$, for $i\not = j \in [k]$, is
\begin{align*}
 \letter{\mathbf{l}, \bar{e},i,j}^{\edge(v_{i,p_i}, v_{j,p_{j}})} 
 \letter{\mathbf{l} ',\bar{\mathbf{e}},i,j}^{|E(i,j)|-\edge(v_{i,p_i}, v_{j,p_{j}})} \letter{\mathbf{l}, i, j}^{|V(i)|-p_i+1}.
\end{align*}
We note that $y$ meets the census requirement $c$. Moreover, the computation of $M$ on input $x$, which selects (as defined in the first part of the proof) exactly the vertices and edges
of the multicolored clique $\set{v_{1,p_1},v_{2,p_2}, \hdots, v_{k,p_k}}$, outputs $y$. Thus $(M=(S,s_0,\Gamma,\Sigma,T), x,c)$ is a \textsc{Yes}-instance for \sGivenWordMM.
\qed \end{proof}

\noindent
The theorem holds if we restrict $\epsilon \notin \Gamma \cup \Sigma$. Indeed, $\epsilon \notin \Gamma$ in the target instance, and one can add a new letter $e$ to $\Sigma$, which replaces $\epsilon$ and has census requirement $c(e) = |x|-\sum_{i,j\in [k],i\ne j} (c(\letter{\mathbf{l},i,j}) + c(\letter{\mathbf{l},\bar{\mathbf{e}},i,j})$. This instance is equivalent since the modified $M$ outputs one letter for each letter in $x$.

\section{Applications}
\label{sec:applications}

In this section we sketch two examples that illustrate how number-of-numbers parameterized problems
may reduce to census problems about Mealy machines, parameterized by the size of the
machine. For another application, see \cite{FellowsLMMRS09}.
\smallskip

\noindent
{\bf Example 1: Heat-Sensitive Scheduling.}
In a recent paper Chrobak et al. \cite{ChrobakDHR08} introduced a model for the issue of temperature-aware
task scheduling for microprocessor systems.  The motivation is that different jobs with the same time requirements may generate different heat loads, and it may be important to
schedule the jobs so that some temperature threshold is not breached.

In the model, the input consists of a set of jobs that are all assumed to be of unit length,
with each job assigned a numerical heat level.  If at time $t$ the processor temperature is
$T_{t}$, and if the next job that is scheduled has heat level $H$, then the processor  temperature at time $t+1$ is 
$$ T_{t+1} = (T_{t} + H)/2 $$
It is also allowed that perhaps no job is scheduled for time $t+1$ (that is,
{\it idle time} is scheduled), in which case $H=0$ in the above calculation of the updated temperature.

The relevant decision problem is whether all of the jobs can be scheduled, meeting a
specified deadline, in such a way that a given temperature threshold is never exceeded.  This problem has been shown to be NP-hard \cite{ChrobakDHR08} by a reduction from {\sc 3-Dimensional Matching}.  An image instance of the reduction, however, involves arbitrarily many distinct heat levels asymptotically
close to $H=2$, for a temperature threshold of 1.

In the spirit of the ``deconstruction of hardness proofs'' advocated by Komusie\-wicz et al.
\cite{KomusiewiczNU09} (see also \cite{BetzlerFGNR09,Niedermeier10}), one might regard this problem as ripe for parameterization by the
number of numbers, for example (scaling appropriately), 
a model based on $2k$ equally-spaced heat levels and a
temperature threshold of $k$.  Furthermore, if the heat levels of the jobs are only roughly
classified in this way, it also makes sense to treat the temperature transition model similarly, as:
$$ T_{t+1} = \lceil{ (T_{t} +H)/2} \rceil $$

The input to the problem can now be viewed equivalently as a census of how many jobs there are
for each of the $2k+1$ heat levels, with the available potential units of idle time 
allowed to meet the deadline treated as ``jobs'' for which $H=0$.  Because of the ceiling function modeling the temperature transition, the problem now immediately reduces to \sExistsWordMM, for a machine on
$k+1$ states (that represent the temperature of the processor) and an alphabet of size at most $2k+1$.  By Theorem \ref{thm:mealyfpt}, the problem is \fpt.
\smallskip

\noindent
{\bf Example 2: A Problem in Computational Chemistry.}  The parameterized problem of {\sc Weighted Splits Reconstruction for Paths} that arises in computational chemistry \cite{GaspersLSS10} reduces to a special case of \sGivenWordMM. The input to the problem is obtained from time-series spectrographic data concerning molecular weights.  The problem as
defined in \cite{GaspersLSS10} is equivalent to the following two-processor scheduling problem.
The input consists of
\begin{itemize}
\item a sequence $x$ of positive integer {\it time gaps} taken from a set of positive integers
$\Gamma$, and
\item a census requirement $c$ on a set of positive integers $\Sigma$ of {\it job lengths}.
\end{itemize}
%
%
The question is whether there is a ``winning play'' for the following one-person 
two-processor scheduling game.  At each step, first, {\it Nature} plays the next positive integer ``gap'' of the sequence of time gaps $x$ --- this establishes the next
{\it immediate deadline}.  Second, the {\it Player} responds by scheduling on one of the two
processors, a job that begins at the last stop-time on that processor, and ends at the
immediate deadline.  The {\it Player} wins if there is a sequence of
plays (against $x$) that meets the census requirement $c$ on job lengths.  Fig. \ref{fig:splits} illustrates such a game.


\begin{figure}[tbh]
\begin{center}
\scalebox{1.2}{
 \begin{tabular*}{0.5\textwidth}{@{\extracolsep{\fill}} r | r r r r r r r}
 Processor 1 & 4 & & 3 & & & 3 &\\
 Processor 2 & & 5 & & 3 & 1 & & 5\\
 \hline
 $x=$ & 4 & 1 & 2 & 1 & 1 & 1 & 4\\
 \end{tabular*}
}
\end{center}
\caption{\label{fig:splits} A winning game for the census: 1 (1),  3 (3),  4 (1),  5 (2)}
\end{figure}

This problem easily reduces to a special case of \sGivenWordMM.
Whether this special case is also $W[1]$-hard remains open.

\section{Concluding Remarks}

The practical world of computing is full of computational problems where inputs are
``weighted'' in a realistic model --- weighted graphs provide a simple example relevant
to many applications.  Here we have begun to explore parameterizing on the
{\it numbers of numbers} as a way of mitigating computational complexity
for problems that are numerically structured.  One might view some of the impulse here
as {\it moving approximation issues into the modeling}, as illustrated by Example 1 in 
Section \ref{sec:applications}.  We believe this line of attack may be widely applicable.

To date, there has been little attention to parameterized complexity issues in the
context of cryptography, control theory, and other numerically structered areas of application.
Number of numbers parameterization may provide some inroads into these underdeveloped areas.

Our main \FPT\ result, Theorem \ref{thm:mealyfpt}, has a poor worst-case running-time guarantee.
Can this be improved -- at least in important special cases?

\smallskip

\noindent
{\bf Acknowledgment.}  We thank Iyad Kanj for stimulating conversations about this work.

\bibliographystyle{plain}
\bibliography{paramNoN}

\end{document}